# Impact of BaTiO$_3$ nanoparticles on phase transitions and dynamics in nematic liquid crystal – temperature and high pressure studies

S. Starzonek,[a,§] S. J. Rzoska,[a] A. Drozd-Rzoska[a], Krzysztof Czupryński[b] and S. Kralj[c,d]

[a.] *Institute of High Pressure Physics of Polish Academy of Sciences, ul. Sokołowska 29/37, 01-142 Warsaw, Poland*

[b.] *Military University of Technology, Institute of Chemistry, ul. Kaliskiego 2, 00-908 Warsaw, Poland*

[c.] *Condensed Matter Physics Department, Jozef Stefan Institute, Jamova 39, 1000 Ljubljana, Slovenia*

[d.] *Faculty of Natural Sciences and Mathematics, University of Maribor, Koroska 160, 2000 Maribor, Slovenia*

[§]E-mail: starzoneks@unipress.waw.pl

## Abstract

Broadband dielectric spectroscopy (BDS) of pure 5OCB and nanocolloids consisting of 5OCB and paraelectric or ferroelectric BaTiO$_3$ nanoparticles (NPs) was performed on varying pressure or temperature. We found strong impact of NPs on static and dynamic phase behavior. In particular, strongest effects on pretransitional behavior were observed for a relatively low concentration of NPs which we attribute to the NPs-induced disorder. Paramagnetic or ferromagnetic character of NPs did not significantly influence measurements. However, several features measured using temperature or pressure path in nanocolloids where significantly different, contrary to the observed behaviour in pure LC compounds. Dynamical properties were tested using the FDSE (fractional Debye-Stokes-Einstein) relation, yielding the fractional coefficient $S = 0.71$ and $S = 0.3$ for the temperature and pressure path, respectively.

## Introduction

In the last decade liquid crystals (LCs) - nanoparticles (NPs) hybrid composites have become the hot topic in the soft matter and liquid crystals physics[1-4]. This is associated with their extraordinary metamaterial features, resulted from the mutually beneficial combination of unique semi-ordered fluids and solid nanoparticles. Nanoparticles can influence the local mesophase order in liquid crystals and parallel, the LC matrix can facilitate NPs arrangements[5-7]. This is supplemented by a set of specific features related to material origins' of NPs: metallic, semiconductor or dielectric. All these can create from LC+NP hybrid composites an exceptional kind of system for new generations of optoelectronic devices[8,9]. However, the fundamental evidence regarding phase transitions is still surprisingly limited, despite the fact that adding of nanoparticles can notably change the LC host system, leading to the emergence of new meta-LC systems. Nanoparticles can act as the moderator of local molecular properties of the liquid crystalline host[4-7]. On the other hand, the local field or geometrical hindrances associated with NPs could influence the local LC symmetry leading to the shift of phase transition temperatures or the decrease of the switching voltage of structural transformations[5].

For the physics of liquid crystals particularly important are forms and impacts of pretransitional effects, their metrics and the exceptional complex dynamics[11]. However, evidences related to these issues for LCs+NPs composites are still very limited. In fact, the first experimental results focusing on critical exponents, the range of critical effects and the value of the discontinuity for weakly discontinuous phase transitions has reported only recently[1,4]. This was the case of dielectric constant behavior in $BaTiO_3$ nanoparticles (non–ferroelectric) – dodecylcyanobiphenyl (12CB, isotropic – smectic A – solid mesomorphism) nanocolloid[1].

This reports presents broad band dielectric spectroscopy (BDS) results for 4'-pentyloxy-4-biphenylcarbonitrile (5OCB) + $BaTiO_3$ nanoparticles (both ferroelectric and non-ferroelectric), yielding insight into dielectric constant and dynamics. Results are both for the temperature and pressure path. It is worth recalling that cooling/heating changes the activation energy whereas compressing influences mainly the free volume changes.

## Experimental

5OCB (4'-pentyloxy-4-biphenylcarbonitrile) belongs to the group of the most 'classical' LC compounds, regarding both fundamentals and applications in displays technology. It is characterized by the following mesomorphism: Isotropic (I) – 68 ºC – Nematic (N) – 53 ºC – Crystal. 5OCB has a strong dipole moment (ca. 4D) at one of the termini of the molecule and approximately parallel to its long axis. The high-purity sample was synthesized in Warsaw Military Technical University (Poland). BaTiO$_3$ nanoparticles (NPs) were purchased from Research Nanomaterials (USA). All of the BaTiO$_3$ phases exhibit ferroelectricity except the cubic phase: this research is associated with the paraelectric cubic phase NPs ($2r = 50$ nm) and ferroelectric tetragonal phase ($2r = 200$ nm).

**Table 1.** Parameters describing pretransitional effects in 5OCB + BaTiO$_3$ NPs nanocolloids in Figure 1. via Eq. (1).

| Sample | NPs size/phase | $T^C$ (°C) | $\varepsilon^*$ | $T^*$ (°C) | $a^*$ | $A^*$ | $\alpha$ |
|---|---|---|---|---|---|---|---|
| 5OCB | – | 69.08$_5$ | 11.369 | 65.323 | -0.0347 | 0.1853 | 0.5 |
| 5OCB+0.1% NPs | 50 nm/cubic | 68.42$_1$ | 10.930 | 68.100 | -0.0243 | 0.1088 | 0.5 |
| 5OCB+0.5% NPs | 50 nm/cubic | 68.82$_5$ | 11.420 | 68.329 | -0.0258 | 0.1244 | 0.5 |
| 5OCB+0.05% NPs | 200 nm/tetragonal | 68.01$_7$ | 10.941 | 61.528 | -0.0515 | 0.3519 | 0.5 |
| 5OCB+0.2% NPs | 200 nm/tetragonal | 68.92$_0$ | 11.066 | 66.996 | -0.0416 | 0.2279 | 0.5 |

LC samples were degassed immediately prior to measurements. Measurements were carried out using the broadband dielectric spectrometer (BDS) Novocontrol, supported by the Quattro temperature controlling system. This enabled 5 digits resolution in dielectric measurements and $\Delta T = \pm 0.02$ °C temperature stabilization.

Mixtures of 5OCB and NPs were sonicated with frequency $f = 42$ kHz for few hours in the isotropic phase until obtaining the homogenous mixture for testing. No sedimentation for at least 24 hours was observed, hence the final colloid did not contained any additional stabilizing agent. Samples were placed in the measurement capacitor made from Invar, with $d = 0.2$ mm gap and diameter $2r = 20$ mm. The quartz ring was used as the spacer. This enable observation of the interior of the capacitor. The latter and the macroscopic gap of the capacitor made it possible to avoid bubbles, distorting results. For each concentration of nanoparticles at least three series of measurements were carried out. The set up for high pressure measurements is shown in Ref. 13: pressure was monitored via two tensometric meters and temperature via the copper-constantan thermocouple within the pressure chamber.

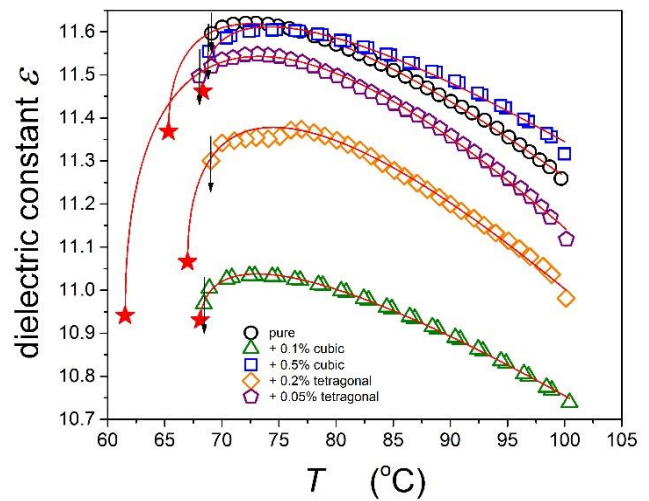

**Figure 1.** The temperature evolution of dielectric constant in pure and BaTiO$_3$ doped liquid crystalline 5OCB (isotropic phase). The red stars denote the extrapolated continuous phase transition temperature and the black arrow the 'real' discontinuous phase transition. The distance between 'stars' and 'arrows' shows the value of the discontinuity of the isotropic – mesophase transition.

## Results and discussion

The evolution of dielectric constant enables getting insight into dominant way of dipole – dipole arrangement in the given system. For the predominantly parallel ordering the value of dielectric constant decreases with rising temperature. When the antiparallel collocation prevails dielectric constant decreases on cooling[12-16]. Such behavior occurs for pretransitional effects in the isotropic phase of 5OCB + NPs (BaTiO3) nanocolloid (Figure 1.). The change of $\varepsilon(T)$ behavior on approaching the mesophase is caused by the growing up number of 5OCB molecules within prenematics fluctuations, for which the equivalence of $\vec{n}$ and $-\vec{n}$ directors is the key feature[5-7]. In each tested case the pretransitional effect can be well portrayed by the same dependence as in 'pure' LC[12-14]:

$$\varepsilon(T) = \varepsilon^* + a(T - T^*) + A(T - T^*)^\phi, T > T_{IM} \quad (1)$$

where $T^* = T_{IM} - \Delta T$ is the extrapolated temperature of the continuous phase transition, $\Delta T$ is the measure of the discontinuity of the isotropic – LC mesophase transition and the exponent $\phi = 1 - \alpha$, $\alpha$ is the exponent of the specific heat. $T_{IM}$ is the isotropic – LC mesophase discontinuous phase transition temperature (clearing temperature).

The parallel of such behavior was reported for the pressure path of approaching the mesophase (nematic, smectic)[4,12]:

$$\varepsilon(P) = \varepsilon^* + b(P^* - P) + B(P^* - P)^\phi, P < P_{IM} \quad (2)$$

where $P^* = P_{IM} + \Delta P$ is the extrapolated temperature of the continuous phase transition, $\Delta P$ is the pressure metric of the discontinuity of the isotropic – LC mesophase transition and the exponent $\phi = 1 - \alpha$, $\alpha$ is the exponent of the specific heat.

The addition of NPs seems to have no impact on the form of parameterization of the pretransitional effects (Figure 1. and Table 1.) but it has the strong influence on the total value of dielectric constant and the discontinuity of the transitions. Adding of the small amount of NPs ($x \approx 0.05\%$) decreases dielectric constant but for higher concentrations the value of dielectric constant notably increase. One can speculate that for very small amounts of NPs the local arrangement of 5OCB is distorted but when increasing the amount of NPs it seems to be strengthened. It is notable that the concentration of NPs has a very strong impact on $\Delta T$ as visible in Figure1 and Table I: in fact for the concentration $x \approx 0.05\%$ of NPs the phase transition approaches to the continuous pattern. Note that similar nonmonotonic temperature behavior has been observed[20-23] in mixtures of different LCs and aerosil NPs. Such behavior is expected when a kind of quenched random field-type disorder is present, what is presented in more detail in the Appendix.

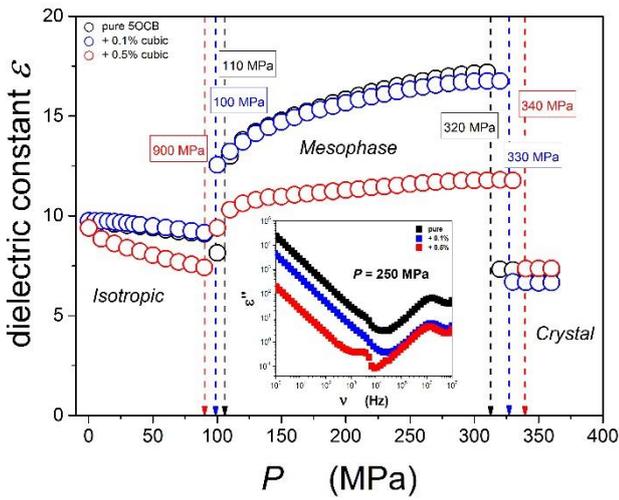 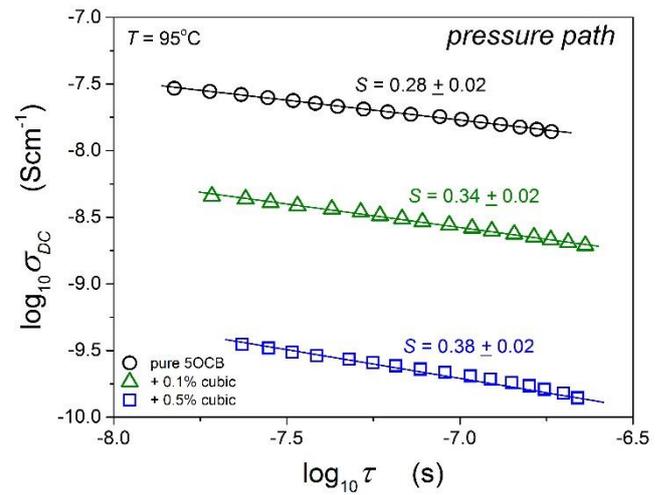

**Figure 2.** The isothermal (T = 95 °C) pressure evolution of dielectric constant in 5OCB and 5OCB + NPs (BaTiO3, paraelectric cubic phase) systems. The insert shows the example of dielectric loss spectra for P = 250 MPa.

**Figure 3.** The plot focused on the translational - orientational coupling ($S = 1$) an decoupling ($S < 1$) via FDSE eq. (1): the behavior in the nematic and nematic – NPs doped mesophases, the pressure path. For the temperature scan the $S$ parameter equals $0.71 \pm 0.02$ in pure liquid crystal.

In the nematic phase the non-oriented sample were monitored, detected the predominantly spontaneous parallel arrangement of permanent dipole moment coupled with 5OCB molecules: it was notably weaker for the highest tested concentration of NPs (Figure 2). When compressing finally the solid phase is reached: there is some impact of nanoparticles here, but much weaker that in fluid phase[4]. Finally, it is notable that for the temperature scan (Figure1) the impact of pressure on the clearing temperature is almost negligible. For the pressure scan the clearing pressure notably decreases with rising concentration of NPs. The question arises if the notable difference for the temperature and pressure behavior can be related to the fact that on compressing nanoparticles can have a preferable locations, associated with 'free volume gaps'[4, 12-16].

Subsequent results reported below focuses on dynamic issues, namely the evolution of dielectric relaxation time and the decoupling between translational and orientational degrees of freedom. The primary relaxation times are determined from determined from peak frequencies as dielectric loss curves, which examples are shown in the inset in Figure 2 as $\tau = 1/2\pi f_{peak}$ [4,17,18].

One can indicate that the inset in Figure 2. shows that the addition of BaTiO$_3$ has very weak impact on the distribution of relaxation times, which metrics are slopes of lines, in the $\log_{10}\varepsilon$ vs. $\log_{10}f$ plot, for $f > f_{peak}$ and $f < f_{peak}$. However, there are other unique features induces by adding NPs: (i) an additional relaxation process in the low frequency region dominated by conductivity emerges, (ii) the range of the region where $(\log_{10}\varepsilon)/(\log_{10}f) = const$ for $f < f_{peak}$ increases from 2 decades (5OCB) to 3 decades (5OCB+0.5%NPs), (iii) adding of NPs decreases by ca. decade the values of $\varepsilon''_{peak}$, which is the rough metric of the energy associated with the reorientation of dipoles; for the low frequency region where $\varepsilon''(f) \propto 1/f$ the distance between 5OCB and 5OCB+NPs case can differ even by 2 decades what suggest a similar distance between relaxation times associated with translational process[19].

The last comment indicated of the comparison of translational and orientational process, which for 'classical' systems are expressed via Debye – Stokes – Einstein (DSE) or its fractional counterpart, namely:

$$\sigma(T,P)[\tau(T,P)]^S = const, \qquad (3)$$

where $\sigma = \sigma_{DC} = 2\pi f \varepsilon_0 \varepsilon''$ for the low frequency part of spectra presented in the inset in Figure 2. The DSE relations ($S = 1$, translational – orientational coupling) converts into the fractional Debye – Stokes – Einstein (FDSE) relation for $S \neq 1$ [4,17,19].

Results of such analysis, in the nematic and nematic-doped mesophase are shown in Figure 3. For 'pure' 5OCB there is a very strong decoupling what can be associated with predominantly parallel ordering of rod-like 5OCB molecules. This decoupling notably weakens when adding nanoparticles, what can be associated with distorting of the parallel arrangement of 5OCB molecules by nanoparticles.

Figure 3. shows results of the similar test focused on the validity of the DSE/FDSE behavior, but for the temperature scan under atmospheric pressure. The most striking feature is not only the occurrence of the FDSE behavior also in this case, but the fact that the translational – orientational decoupling is twice stronger for the temperature

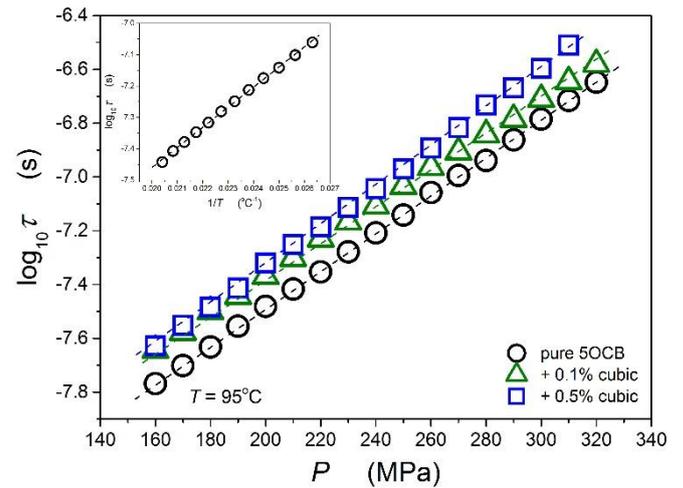

**Figure 4.** The pressure behavior dependence of dielectric relaxation times in 5OCB and 5OCB+NPs (BaTO$_3$) nanocolloids. The insert presents the Arrhenius plot showing the behavior of the temperature evolution of the primary dielectric relaxation time, pure 5OCB.

scan than for the pressure scan. This suggest that for clearly rod-like molecules the general relation (3) valid for low – molecular weight liquids (like glycerol, for instances) should be in the given case replaced by two dependencies[19]:

$$\sigma(T)[\tau(T)]^{S_T} = const \text{ and } \sigma(P)[\tau(P)]^{S_P} = const. \quad (4)$$

All above also may indicate that for rod-like molecules compressing promotes the parallel ordering of rod like molecules.   Figure 4. shows the temperature (a) and pressure (b) evolution of primary relaxation times.  The applied scale shows the clear validity of the simple Arrhenius behavior:

$$\tau(T) = \tau_0 \exp\left(\frac{\Delta E_a}{RT}\right), \text{ for } P = const \quad (5)$$

and its pressure counterpart, the Barus dependence:

$$\tau(P) = \tau_0 \exp\left(\frac{\Delta V_a}{RT}\right), \text{ for } T = const, \quad (6)$$

where $\Delta E_a$ and $\Delta V_a$ stand for the activation energy and volume, respectively.

The value of the activation energy $\Delta E_a$ for pure liquid crystal average ca. 68 kJmol$^{-1}$. For the others sample calculating this value was not possible due to occurring the dielectric relaxation peaks in high-frequency region. The high pressure study give values of activation volume: $\Delta V_a = 9.32$ cm$^3$mol$^{-1}$, $\Delta V_a = 8.78$ cm$^3$mol$^{-1}$, $\Delta V_a = 9.99$ cm$^3$mol$^{-1}$, for pure 5OCB and nanocolloids respectively.

When discussing this issue it is worth recalling that for similar studies in 5CB the clearly visible super-Arrhenius relations was reported, at least for the temperature behavior. The case of the clear Arrhenius/Barus  behavior for 5OCB may be related to the fact that 5OCB molecule is ca. 20 % longer that 5CB[12-14].

## Conclusions

Results presented above revealed a set of  unique features appearing in dielectric static and dynamic properties when adding nanoparticles to a rod-like liquid crystalline host:  in the given case 5OCB and BaTiO3 NPs. First, the ferroelectricity of NPs did not have significant influence on results. Second, the amount of NPs can have a very strong impact on such basic property, regarding fundamentals, as the discontinuity of the phase transition.  Notable is also the fact that patterns observed for temperature and pressure path where different, what in fact could not be revealed for pure LC compounds tested so far.  From both  the application and fundamental point of view important are the strong influence of NPs on the value of dielectric constant, activation energy and FDSE decoupling.

## Appendix: Phase and structural liquid crystal behavior

In the following we illustrate main impacts of NPs on LC behaviour in a nanocolloidal sample. We consider mixtures of LC and immersed identical nanoparticles. Quantities characterizing the samples at the mesoscopic level are the volume concentration field of nanoparticles $c$ and the nematic tensor order parameter $\underline{Q}$ describing a local LC orientational ordering.

Spatially averaged concentration for homogeneously distributed NPs can be expressed as $\bar{c} = N_{NP} v_{NP}/V$. Here $N_{NP}$ counts number of NPs, $v_{NP}$ describes volume of a nanoparticle, $V$ stands for the volume of the sample, and the overbar $\overline{(\dots)}$ marks spatial averaging. The corresponding volume concentration of LC molecules in the mixture is $1 - \bar{c}$. We consider cases $\bar{c} \ll 1$, hence $V \cong V_{LC}$ and $V_{LC}$ stands for the volume occupied by LC molecules. For simplicity we limit to spherical NPs. Their volume and surface equals $v_{NP} = 4\pi r^3/3$ and $a_{NP} = 4\pi r^2$, respectively, where $r$ stands for the NP radius.

We consider thermotropic LCs that exhibit in bulk the isotropic-nematic phase transition at the critical temperature $T_{IM}$ on varying the temperature $T$. At the mesoscopic level the orientational LC order is described with the nematic tensor order parameter $\underline{Q}$. In the uniaxial limit it is commonly expressed as $\underline{Q} = s(\vec{n} \otimes \vec{n} - \underline{I}/3)$. The nematic director $\vec{n}$ identifies the local uniaxial ordering direction, the uniaxial order parameter $s$ measures extent of fluctuations about this direction, and $\underline{I}$ stands for the Identity tensor.

### I. Dilution regime

If NPs are negligible coupled with nematic ordering, they effectively acts as diluting agents, commonly referred to as *impurities*. For NPs of our interest this is realized if the condition $r/d_e < 1$ is fulfilled, where $d_e$ stands for the surface anchoring extrapolation length. It is defined as $d_e \approx K/W$, where $K$ stands for the representative elastic constant and the constant $W$ measures the anchoring strength. For typical LCs it holds $K \sim 10^{-11}$ J/m$^3$ and for a moderate anchoring strength $W \sim 10^{-4}$ J/m$^2$ one gets $d_e \approx 100$ nm.

In the limit $r/d_e \ll 1$ the general thermodynamics well describes resulting behaviour of a mixture. Presence of *impurities* depresses the phase transition temperature $T_{IM}^{(m)}$, where the superscript *(m)* denotes the mixture. It roughly holds that $T_{IM}^{(m)} \approx RT_{IM}^2(\bar{c}_I - \bar{c}_N)/L_{IN}$, where $R$ is the ideal gas constant, $L_{IN}$ is the molar latent heat of the pure *I-N* transition, and $\{\bar{c}_I, \bar{c}_N\}$ label average concentrations of *impurities* in the coexisting nematic ($\bar{c} = \bar{c}_N$) and isotropic ($\bar{c} = \bar{c}_I$) phase, where $\bar{c}_I > \bar{c}_N$. Furthermore, *impurities* enhance the temperature window $\Delta T \propto \bar{c}_I - \bar{c}_N$ of the isotropic-nematic phase coexistence.

## II. Coupling regime

For cases $r/d_e \geq 1$ nanoparticles are sufficiently strong coupled with the nematic ordering to directly influence the local director field. To illustrate most important effects we express the free energy of the soft-nanocomposite as

$$F = \iiint (f_m + f_c + f_e + f_i \delta(\vec{r} - \vec{r}_i)) d^3\vec{r}, \qquad (A1)$$

where $\delta$ stands for the delta function, $\vec{r}_i$ locates nanoparticle-LC interfaces and the integral runs over the LC volume. The mixing ($f_m$), condensation ($f_c$), elastic ($f_e$) and interfacial ($f_i$) free energy densities are given by:

$$f_m = \frac{k_B T}{v_{LC}}(1-c)\ln(1-c) + \frac{k_B T}{v_{NP}} c \ln c + \kappa c (1-c), \qquad (A2a)$$

$$f_c = \frac{3}{2} A_0 (T - T^*) Tr \underline{Q}^2 - \frac{9}{2} B Tr \underline{Q}^3 + \frac{9}{4} C \left(Tr \underline{Q}^2\right)^2, \qquad (A2b)$$

$$f_e = L Tr \left(\nabla \underline{Q}\right)^2, \qquad (A2c)$$

$$f_i = -\frac{3}{2} w_1 \vec{v} \cdot \underline{Q} \vec{v} + \frac{9}{4} w_2 \left(\vec{v} \cdot \underline{Q} \vec{v}\right)^2, \qquad (A2d)$$

where *Tr* stands for the trace operator. We took into account only the most essential terms which are needed to demonstrate qualitatively different phenomena of our interest. Numerical factors in Eq.(A2b) and Eq.(A2d) are introduced for latter convenience.

The term $f_m$ describes the isotropic mixing for the two components within the Flory theory, $k_B$ is the Boltzmann constant, $v_{LC}$ measures the volume of a LC molecule, and $\kappa$ stands for the Flory-Huggins parameter. Note that this parameter promotes phase separation if $\kappa > 0$. Namely, it is minimized for $c=0$ or $c=1$, corresponding to "pure" cases.

The condensation term $f_c$ enforces degree of orientational order. The quantities $A_0$, $B$, $C$ are material constants and $T^*$ describes the supercooling temperature. In bulk equilibrium the phase transition temperature equals to $T_{IM} = T^* + B^2/(4A_0 C)$ and below $T_{IM}$ the corresponding equilibrium degree of uniaxial ordering is given by $s_{eq} = s_0\left(3 + \sqrt{9 - 8(T - T^*)/(T_{IM} - T^*)}\right)/4$, $s_0 = s_{eq}(T_{IM}) = B/(2C)$.

The elastic term $f_e$ is expressed in a single elastic constant approximation and $L>0$ is the representative bare (i.e. independent of $T$) elastic constant. This term enforces a spatially homogeneous structure.

The term $f_i$ describes conditions at a LC-NP interface. We included only some symmetry allowed terms which play important role in our study. The quantities $w_1$ and $w_2$ describe bare surface anchoring strengths and $\vec{v}$ stands for the local normal vector of an interface.

For convenience we introduce the following characteristic lengths of the system which we express at $T=T_{IM}$: $\xi_{IN}^2 = L/(A_0(T_{IM} - T^*))$, $d_1 = w_1 s_0/(L s_0^2)$, $d_2 = w_2/L$. Here $\xi_{IN}$ estimates the uniaxial order parameter correlation length, and $\{d_1, d_2\}$ stand for surface extrapolation lengths.

### II.1 Phase separation

We next estimate the impact of phase separation on critical behaviour in which an orientational ordering is established. To identify main parameters that can trigger phase separation we describe the system by spatially averaged values $\bar{s}$ and $\bar{c}$. In a phase separation process regions displaying different values of $\bar{s}$ and $\bar{c}$ are formed. To get insight into key mechanism we expressed the average free energy density as $\bar{f} \sim \bar{f}_m + \bar{f}_c + \bar{f}_e + \bar{f}_i$ where

$$\bar{f}_m \sim \frac{4k_B T}{v_{LC}}(1 - \bar{c})\ln(1 - \bar{c}) + \frac{4k_B T}{v_{NP}}\bar{c}\ln\bar{c} + \kappa\bar{c}(1 - \bar{c}), \qquad \text{(A3a)}$$

$$\bar{f}_c = (1 - \bar{c})\left(A_0(T - T^*)\bar{s}^2 - B\bar{s}^3 + C\bar{s}^4\right), \qquad \text{(A3b)}$$

$$\bar{f}_e \sim (1 - \bar{c})\frac{L\bar{s}^2}{\xi_d^2}, \qquad \text{((A3c)}$$

$$\bar{f}_i \sim \bar{c}(1 - \bar{c})(-w_1\bar{s} + w_2\bar{s}^2)\frac{r}{4}. \qquad \text{(A3d)}$$

Note that in general $T^*$ is a function of $\bar{c}$. In the simplest case, where the interaction between LC and nanoparticles is negligible, it holds that $T^* = T_0 - \lambda\bar{c}$. Here $T_0>0$ and $\lambda>0$ are independent of $\bar{c}$. Such dependence reflects relatively weaker interactions among LC molecules due to a presence of NPs. The average domain length $\xi_d$ estimates typical elastic distortions in LC orientational order. Furthermore, the average contribution at LC-particle interfaces is proportional with $\bar{c}(1 - \bar{c})$, because this free energy term is absent in the limits $\bar{c} = 0$ and $\bar{c} = 1$.

To understand phase separation tendencies of our system it is convenient to introduce the effective Flory-Huggins parameter $\kappa_{eff}$ as the coefficient weighting the contribution proportional to $\bar{c}(1-\bar{c})$ in $\bar{f}$, i.e.

$$\kappa_{eff} = \kappa + A_0 \lambda \bar{s}^2 - \frac{a_{NP}}{v_{NP}} w_1 \bar{s} + \frac{a_{NP}}{v_{NP}} w_2 \bar{s}^2. \qquad (A4)$$

If $\kappa_{eff}$ is larger than its critical value $\kappa_c > 0$ it triggers phase separation. Namely, its contribution in free energy expression is minimal for $\bar{c} = 0$ or $\bar{c} = 1$. For common LCs it holds true $\kappa \ll A_0 \lambda \bar{s}^2$ and $\kappa < \kappa_c$ and we assume that this is realized in samples of our interest. Let us first neglect the surface interaction terms' contributions. Therefore, mixture is homogeneous in absence of orientational ordering and on entering the nematic phase $\kappa_{eff}$ strongly increases due to the established ordering, which very likely triggers phase separation, leading to the coexistence regime. Furthermore, surface interaction could either enhance or suppress the phase separation tendency. In most samples $w_1$ and $w_2$ are positive, and consequently the $w_1$ ($w_2$) contribution tends to suppress (enhance) phase transition tendency. Therefore, surface treatment could strongly affect phase separation tendency.

## II.2 Nematic structural behavior

We next estimate impact of NPs on LC orientational ordering across the phase transition region for a relatively low concentrations $\bar{c}$, neglecting phase separation. We approximate the free energy $\bar{F}$ of the system by the expression

$$\bar{F} = \left( \left( A_0(T - T^*) + \frac{L}{\xi_d^2} \right) \bar{s}^2 - B\bar{s}^3 + C\bar{s}^4 \right) V + \left( -\bar{s} w_1 + \bar{s}^2 w_2 \right) N_{NP} a_{NP}. \qquad (A5)$$

In expressing Eq. (A5) we estimate average NP induced elastic distortions by $\overline{|\nabla \vec{n}|} \sim 1/\xi_d$ where we assumed mono domain-type structure in the nematic director pattern.

For analytical transparency reasons we introduce the scaled order parameters $\tilde{s} = \bar{s}/s_0$, the dimensionless temperature $t = (T - T^*)/(T_{IM} - T^*)$ and dimensionless free energy density $\tilde{f} = \bar{F}/(V A_0 (T_{IM} - T^*) s_0^2)$. In the following we omit the tildes to avoid clutter and it follows

$$f = t^{(eff)} s^2 - 2s^3 + s^4 - \sigma s, \qquad (A6)$$

where $t^{(eff)} = t + \xi_{IN}^2/\xi_d^2 + (w_2/|w_2|)\bar{c} 3\xi_{IN}^2/(d_2 r)$ and $\sigma = (w_1/|w_1|)\bar{c}\xi_{IN}^2/(d_1 r)$.

The phase behavior of the system described by Eq.(A6) is as follows. For $\sigma < \sigma_c \equiv 0.5$ the first order phase transition takes place when the condition $t^{(eff)} = t_c^{(eff)} \equiv 1 + \sigma$ is realized. The corresponding critical temperature shift $\Delta T = T_{IM}^{(m)} - T_{IM}$ reads

$$\frac{\Delta T}{T_{IM} - T^*} = \frac{w_1}{|w_1|} \frac{\bar{c} 3 \xi_{IN}^2}{d_1 r} - \frac{w_2}{|w_2|} \frac{\bar{c} 3 \xi_{IN}^2}{d_2 r} - \frac{\xi_{IN}^2}{\xi_d^2}. \qquad (A7)$$

Therefore, $w_1 > 0$ (surface ordering potential) tends to increases $T_{IM}^{(m)}$, while $w_2 > 0$ (surface disordering potential) and elastic distortions suppress $T_{IM}^{(m)}$. Above the phase transition the isotropic ordering is replaced by finite paranematic degree of ordering. For $\sigma > \sigma_c$ the supercritical regime is entered where on decreasing temperature the average nematic order parameter gradually increases.

### II.3 Impact of disorder

In our experiments one observes decreased values of dielectric strength in the pretransitional regime for relatively low concentrations of NPs. It is very likely that NPs-induced disorder plays dominant role.

To demonstrate its origin we assume that locally NP-LC interface tends to increase nematic ordering (*i.e.* the term weighted by $w_1 > 0$ dominates at the interface). Consequently, NPs nucleate clusters of size $\xi_d$ exhibiting paranematic ordering. In the low concentration limit it holds $\xi_d \approx \xi$. Here $\xi(T)$ stands for the nematic order parameter correlation length, which is maximal at $T_{IM}$, where $\xi(T_{IM}) = \xi_{IN}$. If $l_{NP} \sim (v_{NP}/\bar{c})^{1/3} > \xi_d$, where $l_{NP}$ estimates the average separation among NPs, then bulk-like behaviour is expected.

Note that for $l_{NP} \geq \xi_d$ paranematic preferential alignment in different clusters is weakly correlated. Furthermore, on approaching the phase transition on decreasing *T* fluctuation enabled clusters exhibiting nematic ordering become more abundant, and they tend to expel *impurities* (*i.e.* NPs) into less ordered surrounding region. Consequently, concentration of impurities increases in paranematic regions, lowering a local value of *l*$_{NP}$. Because fluctuations are strong in these regimes one expect relatively broader distribution of $\xi_d$ values with respect to pure sample. In regimes where $l_{NP} \sim \xi_d$ different clusters interact and in general experiences frustration if their average symmetry breaking direction are significantly different, leading to substantial increase in disorder.

On still increasing $\bar{c}$ nearby clusters become more coupled and correlated, which decreases degree of ordering.

Acknowledgements

The research carried out by was supported via the National Centre for Science (NCN, Poland) grant, ref. 2016/21/B/ST3/02203 (SJR, ADR, SS). SK acknowledge the financial support from the Slovenian Research Agency (research core funding No.P1-0099).